\documentclass[aps,onecolumn,preprint,superscriptaddress,nofootinbib]{revtex4}
\usepackage{amsmath,amssymb,color,mathrsfs, graphicx,verbatim,epsfig, bbm, wasysym, pstricks}
\newcommand{\beq}{\begin{equation}}
\newcommand{\eeq}{\end{equation}}
\newcommand{\beqa}{\begin{eqnarray}}
\newcommand{\eeqa}{\end{eqnarray}}
\newcommand{\tr}{{\rm Tr}}

\newcommand{\cc}{{\rm c.c.}}
\newcommand{\cM}{{\cal M}}
\newcommand{\cO}{{\cal O}}

\begin{document}
\begin{flushright}UMD-PP-09-045 
\end{flushright}
\vspace{0.2cm}

\title{On the Thermal History of Calculable Gauge Mediation}

\author{ Andrey Katz}
\email{andrey(at)umd.edu}
\affiliation{Department of Physics, University of Maryland,\\
College Park, MD 20742}

\date{\today}

\begin{abstract}
\noindent
Many messenger models with realistic gaugino 
masses are based on meta-stable vacua.
In this work we study the thermal history of some of these models.
Analyzing  R-symmetric models, we point out
that while some of the known messenger models clearly prefer the supersymmetric
vacuum, there is a vast class of models where the answer depends on the
initial conditions. Along with  the vacuum at the origin, the high temperature
thermal potential  also possesses a local minimum far away from the origin.
This vacuum has no analog at zero temperature.  
The first order phase
transition from this vacuum into the supersymmetric vacuum is 
parametrically suppressed, and the theory, starting from that vacuum, is
likely to evolve to the desired gauge-mediation vacuum. We also comment on
the thermal evolution of models without R-symmetry.
\end{abstract}

\maketitle

\section{Introduction}\label{intro}

It is well-known that low scale supersymmetry (SUSY) breaking
mediated to the Standard Model through gauge
interactions~\cite{Dine:1981gu,Dine:1982zb,AlvarezGaume:1981wy,Dine:1994vc,Dine:1995ag}
can successfully solve the supersymmetric flavor puzzle. Moreover,
models of gauge mediation are often calculable. Thus
gauge-mediation models should be considered as compelling
candidates for the UV completion of the 
Minimal Supersymmetric Standard Model (MSSM). 

It has been suggested a long time ago (see 
e.g.~\cite{Ellis:1982vi}) that the SUSY-breaking 
minimum can be metastable. It was further argued in~\cite{Dasgupta:1996pz}
that the gauge-mediated vacuum is likely to be metastable 
because of the messenger structure. For additional early examples 
of meta-stable supersymmetric models see 
e.g.~\cite{Poppitz:1996fw,Dimopoulos:1997ww,ArkaniHamed:1997fq}. 
Later it was shown in ISS~\cite{Intriligator:2006dd}
that metastable SUSY-breaking minima are indeed generic and can be found,
for example, in massive SQCD in the free magnetic phase.

Some recent developments even more strongly advocate the idea of
meta-stability of the gauge-mediated vacuum. 
It was proven in~\cite{Komargodski:2009jf} that if
gauginos acquire realistic soft masses within a
calculable model, the gauge-mediation vacuum
must be meta-stable within IR effective theory. This feature leads us to
the question of which vacuum of the theory is cosmologically
preferred.

The issue of the thermal evolution of  supersymmetric models with both
supersymmetric and non-supersymmetric vacua was intensively studied
in~\cite{Abel:2006cr,Craig:2006kx,Fischler:2006xh,Abel:2006my} in the
context of ISS. All of
these studies agreed that the ISS minimum is indeed cosmologically preferable.
Moreover, as it was emphasized in~\cite{Abel:2006cr} the theory is likely
to end up in the non-supersymmetric vacuum even if it begins its thermal evolution
from the supersymmetric vacuum. 

Nevertheless ISS-like meta-stability is not the type of meta-stability which is
required in order to get a viable phenomenology.
The ISS vacuum is \emph{stable within IR renormalizable effective theory},
and the
supersymmetric vacuum becomes manifest only if non-renormalizable
corrections are considered.
Hence one cannot get leading order gaugino masses in ISS
even if R-symmetry is maximally broken by some deformation. The absence
of leading order gaugino masses was explicitly shown in the ISS-based
models of direct gauge
mediation~\cite{Csaki:2006wi,Abel:2007jx,Haba:2007rj,Essig:2008kz}. 
In order not to exacerbate the little hierarchy problem we will further 
consider in our analysis only those theories which have leading 
order gaugino masses. 

Thus we are forced to consider models of a rather different structure than
ISS; dynamical examples of models of that type have been introduced e.g. in~\cite{Kitano:2006xg} and more recently 
 in~\cite{Giveon:2009yu}.  
These examples, as well as other
models one can potentially consider,
are particular dynamical realizations of a much
broader class of models of gauge mediation, so called extra-ordinary
gauge mediation (EOGM)~\cite{Cheung:2007es}. EOGM supplies
the  most generic ansatz for any calculable R-symmetric model with
F-term SUSY breaking one can consider.\footnote{As has been
recently discussed in~\cite{Komargodski:2009pc}, D-terms
are only generated from F-terms in dynamical models of SUSY breaking. 
Thus, usually
D-terms are parametrically suppressed and we will not consider
this possibility in our current study.} The thermal history of
the EOGM models is the focus of our current work.

R-symmetry is an important guideline in analyzing the models of
EOGM. Spontaneously broken R-symmetry, even if it is an
approximate accidental symmetry of the IR effective theory, is well
motivated by the Nelson-Seiberg argument~\cite{Nelson:1993nf}
and because it ensures
that leading order soft terms are CP-conserving.
R-symmetry enables clear classification of the EOGM models,
and endows these models with some useful generic properties,
elaborated in~\cite{Cheung:2007es}.

In this work we systematically study the thermal history of
EOGM models. We first concentrate on a special subset of 
EOGM, 
assuming the ``minimal" UV completion. 
Namely we assume that the
dynamics of the gauge mediation messengers is also responsible for SUSY 
breaking and R-symmetry breaking~\cite{Shih:2007av}. 
The results which we obtain
do not apply to the most general EOGM models. Nonetheless, as we show later,
they turn out to be 
helpful in analyzing models, which are more complicated than just 
the ``minimal completion".

Analyzing various EOGM models we stick to the definitions 
of~\cite{Cheung:2007es}, distinguishing three different types of these models. 
Here we briefly describe these three  types, while more precise definitions 
will be given in section~\ref{survey}.
\begin{itemize}
 \item Type~I models. These are models with the full-rank  
messenger mass matrix. Those models are the most straightforward generalization 
of O'Raifeartaigh models. Type~I models necessarily have vanishing leading 
order gaugino masses.
 \item Type~II models. These are models with full-rank coupling matrix between the messengers 
and the pseudomodulus. The simplest possible example of such models is minimal gauge mediation.
 \item Type~III models. Neither mass nor coupling matrices in these models are full-rank
matrices. Those models share some similarities both with type~I and type~II models.   
\end{itemize}

We argue
that type~II models are cosmologically disfavored if the Universe is reheated
above the SUSY-breaking scale.\footnote{Regarding some
similar models, this observation
has recently appeared in~\cite{Lalak:2009wx}. 
Here we are studying these models in more generality.}
Hence, in order to render these models viable one can obtain a new 
stringent bound on the reheat temperature. Interestingly, this bound
goes precisely in the same direction as bounds on the reheat temperature
from gravitino cosmology, if the gravitino is more massive than 
a keV~\cite{Moroi:1993mb,Giudice:1998bp}.

Studying the type~I models we point out that at high temperatures
they might possess two different vacua. One of them is a global
vacuum at the origin of the field space 
and the second vacuum is local and can potentially
emerge
very far away from the global vacuum. While the messengers in the 
local vacuum are at the origin of the field space, the 
pseudomodulus is stabilized far away from its origin.  
This second vacuum  has no
analog in the zero temperature field theory, and
it disappears when the temperature drops below the SUSY-breaking scale.
In spite of the fact that the free energy of this vacuum is larger 
than the free energy
at the origin, the first order phase transition from 
this vacuum into the vacuum
at the origin is strongly parametrically suppressed.\footnote{Some
examples of the type~I models, e.g. the minimal O'Raifeartaigh
model, have been analyzed in~\cite{Craig:2006kx}. While we agree with
their results regarding the minimum near the origin, we point out
that one more thermal minimum might exist, significantly changing 
the analysis of the thermal history of such models.} We also point out a
necessary condition for this vacuum to exist. We show that this condition
holds in type~I models, that it does not hold in type~II models and 
that it might hold
in type~III models.

Turning to the type~III models, we demonstrate that if the
vacuum far away from the origin exists, the thermal history of the Universe
strongly depends on the initial conditions. If it starts the evolution from
the origin, it will inevitably slide into phenomenologically undesirable 
minimum (or
runaway). On the contrary, if the Universe is trapped after inflation in the
vacuum far away from the origin, its transition rate into the vacuum at the
origin is highly suppressed. When this minimum disappears the system slides
to the gauge-mediation minimum.

Our analysis assumes that
the reheating temperature is higher than the messenger scale
and that the Universe cools down adiabatically.
We restrict ourselves in our analysis to
calculable and renormalizable models. We assume that the
K\"ahler potential is canonical at tree level. For the purposes of
the current analysis we also assume that the Standard Model (SM) 
interactions are much weaker than the messenger model interactions; the 
SM interactions are completely neglected in our current study. It 
would be interesting to study the more general setup in the future. 

Our paper is organized as follows. In the second section we survey the thermal
behavior of different types of ``minimally completed" 
EOGM while also reviewing some facts
about thermal field theory which are necessary for our analysis. 
In that section
we show that  a new vacuum far away from the origin may exist in
type~I models and explain intuitively why one should expect these minima
to show up. In the third section we use these results
in order to explain the behavior of some dynamical 
UV completions
of EOGM found in the literature. 
We also comment on the behavior of
some models which are not R-symmetric. Finally, in the fourth section
we conclude. Some technical details are relegated to the appendix.

%%%%%%%%%%%%%%%%%%%%%%%%%%%%%%%%%%%%%%%%%%%%%%%%%%%%%%%%%%%%%%%%%%%%%%%%%%%%%%%
%%%%%%%%%%%%%%%%%%%%%%%%%%%%%%%%%%%%%%%%%%%%%%%%%%%%%%%%%%%%%%%%%%%%%%%%%%%%%%%

\section{Surveying Reheated EOGM models}\label{survey}
In this section we will survey all three types of R-symmetric EOGM models
and analyze the thermal behavior of each. We will closely
follow the notation of~\cite{Cheung:2007es}. Consider the following superpotential
\beq\label{defEOGM}
W  = \left( \lambda_{ij} X +m_{ij} \right) \varphi_i \tilde \varphi_j
-fX \equiv \cM_{ij} \varphi_i \tilde \varphi_j - fX~.
\eeq
$X$ may acquire both supersymmetric and SUSY-breaking
VEVs $X = \langle X \rangle + \theta^2 F$. Note that the scalar
component of $X$ is a pseudomodulus, which is stabilized at
loop level.

Hereafter we will assume that the SUSY-breaking scale $f$ is much 
lower than the messenger scale $m$. This assumption will further 
allow us to consider the couplings in the hidden sector to be  
of order $\cO(1)$ without sacrificing the zero-temperature 
meta-stability, and justify the formal assumption which we made 
$\lambda \gg g_{SM}$.

At finite temperature $T$ the thermal potential of the model is given
by~\cite{Dolan:1973qd}:
\beqa \label{freeneaxct}
V(T) & = & V(T=0) \pm \frac{T^4}{2\pi^2} \int_0^\infty dx x^2 \ln \left( 1 \mp
e^{-\sqrt{x^2 +\frac{m^2(\phi)}{T^2}}} \right)+ \ldots  \\ \nonumber
&\equiv & V(T=0) + V_{1-loop}(T=0) +V_{th}~.
\eeqa
Plus and minus signs stand for bosons and fermions respectively, and  
summation over all degrees of freedom is implied.
The ellipses stand for higher loop  terms. Note that $V(T=0)$ denotes the 
tree-level zero-temperature potential.

Since we are interested in understanding orders of
magnitude of the phase transition temperatures, we will 
use the expansion of the expression~\eqref{freeneaxct} at temperatures
much higher than the masses of the particles:
\beq\label{freeenht}
V_{th} = -\frac{\pi^2}{90} T^4 \left( N_b +\frac{7}{8} N_f \right)+
\frac{T^2}{24} \left( \sum_{bosons} m^2(\phi) +\sum_{fermions}
\frac{m^2(\phi)}{2} \right) +\ldots
\eeq
The contribution of the particles much heaver than the temperature is Boltzmann
suppressed and can be safely neglected.

%%%%%%%%%%%%%%%%%%%%%%%%%%%%%%%%%%%%%%%%%%%%%%%%%%%%%%%%%%%%%%%%%%%%%%%%%%%%%

\subsection{Type II models} \label{typeii}
We first consider the models with $\det m=0$ and $\det \lambda \ne 0$ (the matrices
$m$ and $\lambda$ are defined in~\eqref{defEOGM}).
These models are clearly phenomenologically interesting since unlike
type~I models they provide viable gaugino masses. The reason for this is that
leading-order gaugino masses are proportional to 
\beq
m_{1/2} \propto \partial_X \ln \det \cM~.
\eeq 
On the other hand it was proven in~\cite{Cheung:2007es} that if $m$ 
is a full rank matrix, then R-symmetry necessarily renders the matrix
$\cM$ to be $X$-independent. Thus the models with $\det m=0 $ are of 
special phenomenological interest. 

Before discussing the thermal history of these models let us first
briefly discuss their vacuum structure at 
zero temperature. Since gaugino masses are non-vanishing at  leading
order, the SUSY-breaking vacuum is necessary metastable. This feature
is manifest: since $\det m=0$, we necessarily have messengers whose
masses are given solely by $\lambda X$. For sufficiently small
values of $X$  
these
messengers become tachyonic. Along the moduli
space of $X$ we
have a stable messenger spectrum only for $X>X_{min}$
for some specific value of $X_{min}$, which is of order 
of magnitude $\frac{\sqrt{f}}{\lambda}.$
Following the tachyonic
direction the theory will either be stabilized at some supersymmetric
minimum or will slide down to a runaway.

Now let us turn to the thermal history of such models. To simplify
the analysis we bring the matrix $\lambda$ to the diagonal form
such that the messenger mass matrix is
\beq
\cM_{ij} = \lambda_i \delta_{ij} X + m_{ij}~.
\eeq
It is straightforward now
to calculate the thermal potential of this theory at temperatures
$T \gg m$, where by $m$ we mean a rough messenger scale. 
The resulting expression is
\beq\label{efptypeii}
V_{th} = {\rm const} + \frac{T^2}{4} \left(\sum_{i=1}^N |\lambda_i|^2
|\varphi|^2 +|\lambda_i|^2 |\tilde \varphi|^2  \right) + \frac{T^2}{4}
\tr (\lambda^\dagger \lambda) |X|^2
\eeq
It is also clear in light of the results of  appendix~\ref{proof} that this 
thermal potential must be independent of $m_{ij}$ and $f$.
The global minimum of the finite temperature potential is at the origin of the
field space. 
One can also verify that no other minimum
emerges far away from the origin, and at 
temperatures $T \gg m$ the minimum at the origin of
the field space is unique.

Since the origin is not a stable point of the zero temperature effective
 potential, we should estimate the temperature at which a phase transition
occurs. The messengers acquire tachyonic masses of order $\sqrt{f}$, so
one expects that the second order phase transition will occur at a comparable 
temperature.
More precisely one can estimate for $\lambda \ll 1$ using
expression~\eqref{efptypeii}
\beq\label{critsecord}
T_{cr} \approx 2 \sqrt{\frac{f}{\lambda}}~.
\eeq
Now we analyze the behavior of the minimum at 
$\varphi_i = \tilde \varphi_i =0  $ with $X$ getting a VEV  
at the messenger scale. Once the system is stabilized at this
vacuum, SUSY is spontaneously broken and the MSSM fields get their soft
masses. We will further call this minimum the ``EOGM vacuum''. 
Let $X_{EOGM}$ denote the value of 
$X$ at that minimum at zero temperature.
When the temperature
is well above the messenger scale, the effective potential for $X$ scales
as $V\sim T^2 X^2$ and the EOGM vacuum does not exist. 
Once the temperature
drops below $\lambda X_{EOGM} $  all the messengers near the point
$X=X_{EOGM}$ become heavy (since $\lambda$ is a full rank matrix) and the EOGM
minimum emerges. Nonetheless the number of  thermalized degrees of freedom
in the EOGM minimum is smaller than at the origin. In general, 
one can notice 
that the free energy of the EOGM minimum is bigger than the free energy of the origin
(or, alternatively, the supersymmetric vacuum after the second order
phase transition occurs) \emph{at any temperature}. 

This claim can 
be explained as follow. It is easy to see
from~\eqref{freeenht} that a vacuum with a larger number of light 
degrees of freedom has lower free energy. When the EOGM vacuum is 
formed, the only light supermultiplet in this vacuum is $X$. 
It becomes heavy at a temperature of order 
$\sqrt{\frac{\alpha_\lambda}{4\pi } f}$. On the other hand, the vacuum at 
the origin has more light degrees of freedom: both $X$ and some messenger 
pairs are light (such messengers exist since 
${\rm rank}\ \lambda>{\rm rank}\ m$). Once we get to the temperature of 
order $f$, the free energy of the EOGM vacuum is already governed by the 
temperature independent term, $f^2$, which arises since EOGM vacuum 
breaks SUSY. 
Hence the thermal potential at the EOGM vacuum below the 
temperature~\eqref{critsecord} scales as 
\beq\label{VEOGM}
V_{EOGM} (T^2 < f) =  f^2 - \frac{\pi^2}{24} T^4 +\ldots \approx f^2~. 
\eeq  
Clearly this value is bigger than the thermal potential along the path 
from the origin to the supersymmetric vacuum. The key point is that along
that path the SUSY-breaking terms $f^2$ is absent and the value of the thermal 
potential is governed by the number of the massless degrees of freedom, which 
is necessarily smaller than~\eqref{VEOGM}.     
 Thus,
type~II minimally-completed EOGM models are always cosmologically disfavored
unless the reheat temperature is significantly smaller than the messenger
scale. Therefore in this context we find a new bound on reheat temperature.   

%%%%%%%%%%%%%%%%%%%%%%%%%%%%%%%%%%%%%%%%%%%%%%%%%%%%%%%%%%%%%%%%%%%%%%%%%%%%
\subsection{Type~I models}\label{typei}
In this subsection we discuss the models with $\det m\ne 0$ and
$\det \lambda =0$. Since these models cannot produce reasonable phenomenology,
we will analyze this scenario as a toy model. The results will
be  useful in the analysis of type~III models.

Before analyzing genuine type~I models, consider as an example
the minimal O'Raifeartaigh model (let the fields
$\varphi,\ \tilde \varphi$ be gauge singlets in this example):
\beq\label{typeiW}
W = \lambda X ( \tilde \varphi^2 -f) + m\varphi \tilde \varphi~.
\eeq
It is straightforward to calculate the thermal potential 
for small $X$ in this model
and verify that at sufficiently high temperatures all the fields can be stabilized
at the origin. 

Let us now analyze this model at very large values of $X$, namely in the
range
\beq\label{runawayreg}
X \gg T, \ \ X \gg m~.
\eeq
The key point for understanding this regime is noting that at zero
temperature one finds very light particles in the O'Raifeartaigh models
for $X \gg m$. Integrating out $\tilde \varphi$ in~\eqref{typeiW}
for large $X$ one finds the effective superpotential
\beq
W = -\lambda X f -\frac{m^2}{4 \lambda X}  \varphi^2~.
\eeq
Notice that the mass of $ \varphi$ is suppressed, rather than
enhanced, by $X$. Since the zero-temperature potential for $X$ vanishes
at tree level, the non-thermal mass of the pseudomodulus is also
strongly suppressed, such that both $\varphi$ and $X$ are thermalized in
the regime~\eqref{runawayreg}. The mass of $\varphi$ becomes exactly
zero when $X$ gets to infinity, so we should expect in this range
thermal runaway for $X$.
In order that this regime be reliable we need both inequalities
in~\eqref{runawayreg} to hold. If the temperature exceeds $X$, the field
$\tilde \varphi$ is also thermalized and the runaway behavior is lost.

Now let us ask whether this runaway is stabilized. The thermal potential for
$X$ is 
\beq
V_{th} \sim \frac{m^4 T^2}{|\lambda X|^2}~.
\eeq
 But the zero temperature potential is a monotonically increasing function.
The leading order term of the one-loop effective potential is
\beq\label{oneloopfar}
V_{1-loop} \sim \frac{\alpha_\lambda}{4 \pi} f^2 \ln |X|^2
\eeq
which will inevitably stabilize the runaway.
The mass matrix for the field $\varphi$ is positive definite
in the regime~\eqref{runawayreg}
if $m^2 > f$. Thus this simple
O'Raifeartaigh model, besides  the vacuum at the origin, gets a local
minimum of the thermal potential far away from the origin,
$X\gg T$. More precisely the potential is balanced at 
\beq
X_* \sim \left( \frac{\alpha_\lambda}{4 \pi}\right)^{-1/2} 
\frac{m^2 T}{\lambda f}~. 
\eeq  
The free energy of this local vacuum is bigger than the free
energy of the vacuum at the origin, so we are supposed to check if 
the thermal transition rate from this vacuum to the origin is
suppressed. It turns out that the transition rate for the first
order phase transition to the vacuum at the origin is parametrically
suppressed since $X_* \gg T$. We will justify this statement
in the subsection~\ref{typeiii}, after extending our analysis to broader
class of models.

Now we are ready to generalize this idea to the type~I EOGM models. Consider
for example a model with $N$ messenger pairs and with the following superpotential:
\beq
W = m_{(i)}\varphi_i \tilde \varphi_i+ \lambda_{(i)} X \varphi_i
\tilde \varphi_{i+1} -fX~
\eeq
with $i$ running over the values $1\ldots N$.
At large values of $X$ all of the messengers excluding $\varphi_N$ and
$\tilde \varphi_1$ are supposed to be integrated out. 
After integrating
out the heavy messengers one gets the following superpotential:
\beq
W = -f X + (-)^{(N-1)}
\frac{\det m }{\prod_i \lambda_{(i)} X^{N-1}} \varphi_N
\tilde \varphi_1~.
\eeq
Note that the power of $X$ in this expression can be determined, without any 
calculation, from R-symmetry considerations. 

Now we check
if the messenger mass-squared matrix is positive definite at
zero temperature. The mass-squared matrix for the scalars is given by two
$2\times 2$ diagonal blocks of the form
\beq\label{typeiztmassmat}
M^2 = \left( \begin{array}{cc}
\left(\frac{\det m}{\prod_i \lambda_i} \right)^2 \frac{1}{X^{2N-2}} &
\pm (N-1) \frac{\det m }{\prod_i \lambda_i}\frac{f}{X^N}\\
\pm (N-1) \frac{\det m }{\prod_i \lambda_i}\frac{f}{X^N}&
\left(\frac{\det m}{\prod_i \lambda_i} \right)^2 \frac{1}{X^{2N-2}}
\end{array} \right)~.
\eeq
Since we are discussing the case where $X$ is large, this matrix \emph{is
not necessarily positive definite}.  
This is of course nothing but a 
manifestation of the claim of~\cite{Cheung:2007es} that the moduli space
of the type~I EOGM might become unstable when $X$ exceeds some
$X_{max}$.\footnote{For detailed discussion of this instability
see~\cite{Ferretti:2007ec}.}
Nonetheless, for $N=2$ (or, alternatively, multiple decoupled
species of $N=2$)
the matrix~\eqref{typeiztmassmat} is always positive
definite. 

Consider the thermal potential. Up to $\cO(1)$ numbers the leading terms
of the thermal potential are
\beq
V_{th} \sim \left( \frac{\det m}{\prod_i \lambda_i} \right)^2
\frac{T^2}{|X|^{2N-2}} + \left( \frac{\det m}{\prod_i \lambda_i} \right)^2
\frac{T^2}{|X|^{2N}} \left( |\varphi_1|^2 + |\tilde \varphi_N|^2 \right)~.
\eeq
As expected, we get runaway behavior for $X$, and thermal masses for the light
messengers which are highly suppressed compared to
the naive expectation. The minimum always exists in models
with two messenger pairs; in models with more messenger pairs it might
exist for some definite range of temperatures. In these models
one should check if the matrix~\eqref{typeiztmassmat}
is positive definite for a given choice of parameters.

Now we are ready to estimate the position of the vacuum. We will continue
doing analysis for generic $N$, but one should keep in mind that only for
$N=2$ (or the set of decoupled models of this type) the existence of this vacuum is guaranteed.  
The one-loop effective
potential far away from the origin is given by~\eqref{oneloopfar}.\footnote{Now
this expression may be imprecise since we have different
coefficients $\lambda_i$. Of course the dominant contribution here will
come from the largest value of $\lambda$.}
The minimum of the thermal potential scales as
\beq\label{farawayminimum}
X_* \sim \left( \left( \frac{\alpha_\lambda}{4\pi} \right)^{-1/2}
 \frac{\det m}{\prod_i \lambda_i} \frac{T}{f}\right)^{\frac{1}{N-1}}~.
\eeq
Consider now that we start the thermal evolution in the vacuum far
away from the origin and gradually lower the temperature. The vacuum
will disappear either if $X_* \sim \sqrt[N]{\det m}$
and integrating some of the messengers becomes unjustified, 
or if the messengers at that vacuum
become heavier than the temperature, whatever happens earlier.
The second condition is stronger; the vacuum disappears at temperature: 
\beq\label{lowbound}
T \sim \left( \sqrt{\frac{\alpha_\lambda}{4\pi}} f \right)^{1/2}~,
\eeq
namely right below SUSY-breaking scale. 

Note also that only for $N=2$ this vacuum survives up arbitrarily high
temperatures. If the number of messenger pairs is larger than $2$, we get 
an upper bound on the temperature demanding that matrix~\eqref{typeiztmassmat}
is positive definite. This upper bound reads
\beq\label{upbound}
T < \sqrt{\frac{\alpha_\lambda}{4\pi}} \left( \frac{\det m}{\prod_i \lambda_i}
\frac{1}{f} \right)^{1/(N-2)}~.
\eeq     
Evidently for sufficient separation between the messenger scale 
and the SUSY-breaking scale, the window defined by~\eqref{lowbound}
and~\eqref{upbound} can be very large.

One can generalize these considerations to other messenger models, 
not necessarily
of type~I. Assume that we have several sets of messengers
such that the matrix $\cM$ is block diagonal. We 
note that a necessary (but not sufficient!) condition in order to get
this behavior is to have some subset of messengers which
fulfill the condition
\beq\label{neccond}
 {\rm rank}\ m > {\rm rank }\ \lambda~.
\eeq
If all of our messengers fulfill this condition, we get an EOGM model
of type~I, but in general type~III models can also exhibit this 
behavior.

%%%%%%%%%%%%%%%%%%%%%%%%%%%%%%%%%%%%%%%%%%%%%%%%%%%%%%%%%%%%%%%%%%%%%%%%%
\subsection{Type~III models} \label{typeiii}
Now we turn to the most phenomenologically interesting part of our
analysis. These models have both $\det \lambda=0$ and $\det m =0$
and share the features of both type~I and type~II models.
In particular these models are well motivated since they lead
to non-vanishing gaugino masses at  leading order.
The vacuum
structure of these models at zero temperature is also non-trivial:
they might develop instability both for very small values of the
pseudomodulus (as in the case of type~II models) and for very
big values (as in the case of type~I).

The thermal histories of these models also share similar features with both
 type~I and type~II theories.
It is clear that the thermal potential possesses a global
minimum at the origin. If the Universe starts its evolution from this
minimum, it will undergo a second order phase transition at the
temperature~\eqref{critsecord}, as was described in
subsection~\ref{typeii}, and slide into the supersymmetric vacuum or runaway.

Nonetheless, sliding into the supersymmetric direction is not the only
possibility. Since these models might have some set of
messengers fulfilling the condition~\eqref{neccond}, we can
expect that certain models will also
posses a local thermal vacuum far away from the origin.
Here we show an explicit example
of a model which indeed exhibits this behavior. Moreover, we show that
starting the evolution from this minimum one ends up in the EOGM minimum,
rather than in the supersymmetric one.

Consider a model with following messenger mass matrix:
\beq
\cM = \left( \begin{array}{cccc}
\lambda' X & 0 & 0 & 0\\
0 & m & \lambda X & 0 \\
0 & 0 & M & \lambda X\\
0 & 0 & 0 & m
\end{array} \right)~.
\eeq
This model  was analyzed in~\cite{Cheung:2007es} as an
example of the ``minimal completion" of type~III EOGM. 
It was shown that
for sufficient separation of scales $m$ and $M$ the pseudomodulus
can be stabilized between these two  scales.\footnote{As  explained
in~\cite{Giveon:2009yu}, the separation of masses is a necessary condition
for the radiative stabilization of moduli in this kind of model.}

Consider now this model thermalized when $X$ is sufficiently far away
from the origin. The first pair of messenger decouples,
three other pairs maintain the condition~\eqref{neccond}.
We can expect that the vacuum far away from the origin will show up. In
order to know the exact scaling of this vacuum we just substitute  $N=3$
in the formulas which we derived in the previous subsection. We find 
from~\eqref{farawayminimum}
\beq\label{xstar}
X_* \sim \left( \left( \frac{\alpha_\lambda}{4\pi}\right)^{-1/2} 
\frac{m^2 M}{\lambda^2}
\frac{T}{f} \right)^{1/2}~.
\eeq
As  explained in the previous subsection,
the vacuum far away from the origin for $N>2$
exists
only at certain range of temperatures. In our 
case this range is given by
\beq\label{allrange}
\left( \sqrt{\frac{\alpha_\lambda}{4\pi}}\ f \right)^{1/2}<T <
\sqrt{\frac{\alpha_\lambda}{4\pi}} \frac{m^2 M}{\lambda^2 f}
\eeq 
Once the vacuum dissipates,
the modulus $X$ is at the messenger scale. Substituting the lower 
bound on the temperature from~\eqref{allrange} into~\eqref{xstar}
we explicitly verify that when the vacuum far away from the origin dissipates,
the value of $X_*$ still respects an inequality
$X_* \apprge X_{EOGM}$. 

In order to understand what subsequently happens to the 
system which evolves through the vacuum far away from the origin, we analyze
the thermal potential at the SUSY-breaking temperature. All the messengers are stable 
at the messenger scale, they have masses of order $M$ and the potential rises 
in these directions, such that the pseudomodulus direction $X$ \emph{is the only 
relevant direction for this discussion.} Since at the temperature 
$T \sim \sqrt{f}$ none of the particles besides the  multiplet $X$ are thermalized
in the vicinity of the EOGM vacuum,
the shape of the thermal potential is the same as it is at zero temperature. 
Namely the potential monotonically rises at $X>X_{EOGM}$ approximately as 
$V \sim \log X + \ldots $, and decreases for smaller $X$ towards the local 
SUSY-breaking EOGM minimum. Therefore we conclude that when the vacuum far away from 
the origin dissipates, the system undergoes second order phase transition to the 
EOGM minimum. On the other hand, the supersymmetric minimum is located near the origin.  
Since the shape of the potential is similar to the shape of the zero temperature 
potential, the supersymmetric vacuum is separated from the EOGM minimum by a barrier
with height of order $\sqrt{f}$,  and the distance between two these minima is of 
order $M$. Consequently, a second order phase transition from the vacuum far away from the 
origin into the supersymmetric vacuum is impossible.    

Now let us briefly discuss the hypothetical possibility of a
first order phase transition from the vacuum far from the origin. The
rate for this phase transition is given by
\beq
\Gamma \sim T^4 e^{-\frac{S_3}{T}}~.
\eeq
In our case the potential rises from the origin as a quadratic function
of $X$, up to the values $X\sim T$. At this value the potential for 
$X$ starts falling as a negative power.
The three dimensional bounce action can be approximated through the triangular
approximation~\cite{Duncan:1992ai}:
\beq\label{bonceac}
\frac{S_3}{T} \sim 4\pi \frac{(\delta X)^3}{\sqrt{\delta V}T }
\sim 4\pi \frac{X_*^3}{T^3}~.
\eeq
Since we obtained $X_*\gg T$, this rate can be made arbitrarily small
by appropriate choice of parameters for any temperature when the 
vacuum far away from the origin exists. Note that by the same reason 
the first order phase transition rate from the EOGM minimum at the temperature
$\sqrt{f}$ (or lower) to the supersymmetric vacuum is also parametrically suppressed.
In this last case one should of course substitute $M$ instead of $X_*$ 
into~\eqref{bonceac}.

%%%%%%%%%%%%%%%%%%%%%%%%%%%%%%%%%%%%%%%%%%%%%%%%%%%%%%%%%%%%%%%%%%%%%%%%%
%%%%%%%%%%%%%%%%%%%%%%%%%%%%%%%%%%%%%%%%%%%%%%%%%%%%%%%%%%%%%%%%%%%%%%%%%
\section{Comments on dynamical models}\label{comment}

\subsection{R-symmetric Dynamical and Retrofitted Models}\label{dvr}
In the previous section we  studied  models of messengers
with the ``minimal" completion. It is plausible that at low
energies this is a correct description, while all the
small mass scales are just retrofitted as in~\cite{Dine:2006gm}.
In this case the description of the thermal behavior of these models
is exhaustively explained in the previous section. Nonetheless,
we will also be interested in applying our results to models with
extra matter, which reproduce the EOGM ansatz dynamically.

Consider first the type~III models. Very few dynamical examples
of this kind of model are known, one of them presented, e.g.
in the model of 
``uplifted vacua in SQCD"~\cite{Giveon:2009yu}.\footnote{An earlier
example was given in~\cite{Kitano:2006xg}. The thermal behavior of 
these two models does 
not differ.}
One can wonder if the mechanism of
``rescuing" the meta-stable vacuum through evolution far away from
the origin is still available in this case. 
The answer is negative. The crucial difference between the model presented
in~\cite{Giveon:2009yu} and the ``retrofitted" type~III model is that
in the retrofitted models the masses are ``replaced" by the VEVs
of some gaugino condensates. These condensates are usually formed at very
high temperatures, and at  low temperatures the masses behave as
fundamental parameters.  On the other hand, in dynamical models,
these masses do not emerge before
the thermal vacuum at the origin is destabilized
and the relevant fields acquire VEVs.
This usually happens at relatively low temperatures. Above that
temperature the system evolves as if it did not have any masses at all,
in other words similarly to the type~II models. For example in the
``uplifted vacua model" the masses do not emerge before the Universe
is cooled down below the messenger scale, and hence the mechanism
described in  subsection~\ref{typeiii} for rescuing the EOGM vacuum
is inapplicable. Therefore, a stringent bound on the reheat temperature
of these models applies.

The next question we address is the behavior of (R-symmetric) models
where messengers are added ``by hand'' and they are not part of
the SUSY breaking sector.
To illustrate this kind of model,
consider an R-symmetric model of messengers and SM 
singlets which couple to the
pseudomodulus. If the origin
of some field is destabilized at sufficiently high temperatures, they
must have vanishing R-charge.\footnote{We prove this statement in the
appendix.} Hence, at sufficiently high temperatures $X$
has a minimum at the origin independent of the details of the SM singlets
to which it couples. If the theory starts from this vacuum, either by choice
of initial conditions or because it is the only vacuum of the theory, it
will inevitably cool down to the supersymmetric vacuum. The possibility
of utilizing the rescuing mechanism through the vacuum far away from the origin
strongly depends on the details of the theory. If the rank of the matrix
$m$ exceeds the rank of the matrix $\lambda$ for any subset of messengers or
SM singlets, the vacuum far away from origin might emerge, as discussed
in section~\ref{survey}. As previously noted, the
masses should emerge at sufficiently high temperatures.

Applying these arguments e.g. to the model~\cite{Giveon:2008ne}, which is
R-symmetric and includes both ``hidden sector'' fields 
and messenger fields (which behave as in type~II),
we conclude that the rescuing mechanism through the vacuum far 
away from the origin does not work.
Even though the hidden sector fields perfectly
fulfill the condition~\eqref{neccond}, the masses form only at the
SUSY breaking scale, which is not sufficient to ensure the vacuum far away
from the origin. We note, however, that the interactions between the messengers
and the pseudomodulus in~\cite{Giveon:2008ne} can we extremely suppressed.
While the model with the low scale $M_2$ will favor the supersymmetric 
minimum, unless the reheat temperature is lower than the SUSY-breaking 
scale, the thermal history of the model with high $M_2$ will be largely 
governed by the SM interactions, which are not taken into account in our 
current study. If the SM D-terms indeed prevail, the final answer can change 
since the temperature for the second order phase transition into the EOGM
vacuum can exceed the phase transition temperature in the supersymmetric
direction.\footnote{I am grateful to S.~Abel, J.~Jaeckel and V.~Khoze for 
pointing me out the importance of the SM interactions.}   

%%%%%%%%%%%%%%%%%%%%%%%%%%%%%%%%%%%%%%%%%%%%%%%%%%%%%%%%%%%%%%%%%%%%%%%%%%%%%%
\subsection{Comments on models without R-symmetry}\label{nors}
In this subsection we briefly comment on messenger models  which lack
R-symmetry. As an example we will analyze here models
presented 
in~\cite{Murayama:2006yf,Murayama:2007fe,Aharony:2006my}.\footnote{A 
similar model has been recently introduced in~\cite{Abel:2009ze}. Since
in~\cite{Abel:2009ze} one should take $\lambda\ll 1$, the SM effects may 
become important.}
In these models the messenger sector was coupled to the ISS model,
while the messenger masses were added ``by hand".\footnote{These
masses were further ``retrofitted" in~\cite{Aharony:2006my} but
it does not affect the thermal behavior of the model.}
These models exhibit a peculiar behavior: unlike R-symmetric models
they might be stabilized at high temperatures at a point which is
not a zero of the $\det(\lambda X + m)$.\footnote{Note, that in the 
ISS model $\langle X \rangle =0$.}

Consider for simplicity  the following
 superpotential:
\beq\label{nonRW}
W = \lambda X \varphi_i \tilde \varphi_i +\lambda' X \phi_j \tilde \phi_j
-m\varphi_i \tilde \varphi_i -fX~.
\eeq
The messengers $\varphi, \tilde \varphi$ are charged under the SM model
while $\phi,\ \tilde \phi$ are uncharged.\footnote{Those fields
may be multiplets
of some other symmetry group, but it is irrelevant for our
discussion.} It is easy to see that the thermal potential of the
model~\eqref{nonRW}  up to an overall numerical factor is
\beq
V_{th} \propto T^2\left( N_\phi |\lambda X|^2 +
N_\varphi |\lambda X-m|^2 \right)~.
\eeq
At sufficiently high temperature the pseudomodulus is stabilized at
\beq\label{tvnR}
X = \frac{\lambda N_\varphi m}{\lambda' N_\phi +\lambda N_\varphi}~.
\eeq
Both $X=0$ and $X=m$ are zeros of  $\det(\lambda X-m)$, but
the stabilization is at some intermediate value between 
them.\footnote{As a consequence of appendix~\ref{proof}, an
R-symmetric O'Raifeartaigh model cannot exhibit such behavior
unless the R-charge of $X$ vanishes.}

Now we trace the evolution of this vacuum when the temperature
drops down. Of course, we assume that the scale $m$ is much larger
than the SUSY breaking scale (otherwise even the zero-temperature
meta-stability may be spoiled). At the scale $m$ the vacuum
at~\eqref{tvnR} is destabilized and the second order phase transition occurs.
The zero temperature masses of the fields $\varphi,\ \tilde \varphi$ and
$\phi,\ \tilde \phi $ for $X=X_*$ are $(\lambda X_* - m)$ and $\lambda' X_*$
respectively. Those particles which are more massive will stop affecting
the thermal potential at higher temperatures since their contributions
will be Boltzmann suppressed, and the theory will slide into the direction
where the lightest particles attract it. We can draw the conclusion that
the models of ISS with messengers are not automatically safe;
their thermal history strongly depends on the ratio
$\frac{\lambda N_\varphi}{\lambda' N_\phi}$. If $\lambda N_{\varphi}$
``wins'', the theory will be attracted to the direction of
the supersymmetric minimum, but since these models have no restrictions
on the hidden sector, this potential problem can circumvented by choosing
a big enough magnetic group in the ISS sector.

Now let us consider a more generic picture and try to draw some conclusions.
A thermal potential of an arbitrary messenger model without R-symmetry
is a sum of quadratic functions with the origins sitting at the zeros of
 $\det (\lambda X+m)$. Thus the minimum of the potential is likely
to be somewhere between these zeros. At  zero temperature
each of these zeros, if it is simple,
 is necessarily an attractive point of the potential.
I.e. there is some vicinity of each of these points where the
potential for $X$ attracts it to the zero, independently of whether this
point has tachyonic messengers~\cite{Zoharnote}.\footnote{In a generic model
without R-symmetry we expect all zeros to be simple.}
Namely each of these zeros can govern the thermal evolution, and
an exact answer is highly model-dependent, as illustrated in the
example of~\eqref{nonRW}.

\section{Conclusions and Outlook} \label{cons}
In this paper we studied the thermal behavior of various gauge mediated models,
mostly with, but also without R-symmetries. We showed that at high temperatures
all R-symmetric messenger models possess a minimum at the origin of field
space. Starting from this minimum the models generally evolve to the
supersymmetric vacuum. 
In order to avoid this pitfall  
one should
demand that the reheat temperature is lower than the messenger scale and that
the Universe is trapped in the false vacuum after inflation. Demanding
reheat temperature lower than the messenger scale can be quite a severe
constraint on low-scale gauge-mediated models.

We also showed that this stringent demand of the low reheat temperature
can be circumvented in certain models, where some particles, coupling to
the pseudomodulus maintain the condition~\eqref{neccond}. In this case
an additional minimum can emerge far away from the origin which can drive
the thermal evolution of the Universe. In this case the Universe may end
up in the SUSY-breaking minimum even for relatively high reheat temperature.
Clearly the thermal evolution of models of this class strongly depend
on the initial conditions after inflation. The subject of the initial 
conditions was not studied here and it would be interesting to study
this subject in future works.  

We have presented convincing evidence that the thermal history is more special
if the model is R-symmetric. It would be interesting to study further 
the connection between R-symmetry and the 
structure of the thermal potential.

We also notice that we did not study the effects of the SM gauge interactions,
concentrating solely on the messenger sector analysis. While the messenger
sector effects are extremely important, and the thermal history 
of lots of models can be determined only by analyzing the messenger dynamics,  
it would be interesting
to accomplish this analysis including the SM effects.
Clearly the approximation where the 
SM effects are completely neglected cannot be valid in all models and in
all parts of the parameter space, so it would be important to understand 
in the future, where precisely these effects are important and whether they can
relax some of the constrains on the gauge-mediated models.       

\vskip 0.2in \noindent {\bf Acknowledgements:}\ \ I am indebted
to Z.~Chacko, Amit Giveon, Raman Sundrum, Brock Tweedie 
and especially Zohar Komargodski and David Shih  for
illuminating discussions.  I am also grateful to Steven Abel,
Joerg Jaeckel and Valya Khoze for the communication and discussion 
of the SM effects. The
work was partially supported by NSF under grant PHY-0801323.

\appendix
\section{Properties of the Thermal Potential in R-symmetric Models}\label{proof}
In this appendix we consider a thermalized R-symmetric O'Raifeartaigh
model. We prove that at sufficiently high temperatures
the origin of the field is destabilized only if its R-charge vanishes.
We also show that the field can possess vanishing thermal mass
at the origin only if
there is more than one field in the theory with precisely  the same R-charge.

Consider the most generic renormalizable R-symmetric Wess-Zumino model:
\beq\label{WZW}
W = f_i \phi_i + \frac{m_{ij}}{2} \phi_i \phi_j +\frac{\lambda_{ijk}}{6}
\phi_i \phi_j \phi_k~.
\eeq
The couplings $m_{ij}$ and $\lambda_{ijk}$ are symmetric in
all indices.

Now consider  temperatures much higher than the masses in~\eqref{WZW}
and derive the leading order field dependent thermal potential.
Analyzing the approximation~\eqref{freeenht} we notice an important
feature which significantly simplifies our analysis: the leading order
thermal potential \emph{is merely sensitive to the trace of the mass
squared matrix}. Namely, we will not be be interested either in the
SUSY-breaking terms in the zero temperature potential or in any other
off-diagonal terms in the mass squared matrix.\footnote{To use this property
we need an additional assumption $\sqrt{f} \ll m$. Fortunately it holds in the
vast majority of realistic gauge-mediated models.} 
In order to obtain a
correct leading term thermal potential it is sufficient to take a sum
over the terms in the zero-temperature tree-level potential of
the form
\beq\label{contrterm}
V \supset m_i^2(\phi) \phi_i \phi^*_i~.
\eeq
The quadratic terms in the thermal potential are always positive since they are
formed from the interaction in~\eqref{WZW} . So
the only terms which can cause destabilization in the thermal potential are
either tadpoles or non-diagonal quadratic terms.

In order to get a tadpole in the thermal potential we
should have trilinear term
in the zero temperature potential. The only way to get these interactions
is
\beq
V \supset m^*_{kl} \phi^*_l \lambda_{ijk} \phi_i \phi_j +\cc
\eeq
Since  terms which may give contributions to the thermal potential
should be of the form~\eqref{contrterm}, either index $i$ or index $j$
should be equal to $l$. Namely the field $\phi_j$ will have a tadpole
in the thermal potential only if the terms $\phi_k \phi_l$ and
$\phi_k \phi_l \phi_j$ are both present in the superpotential. This
can happen only if the R-charge of $\phi_j$ is zero.

Another source of thermal instability can potentially
come from non-diagonal
masses in the thermal potential. Such masses emerge from the quartic
terms in the zero-temperature potential:
\beq
V \supset \lambda^*_{ijk} \lambda_{ijl} |\phi_j|^2 \phi^*_k \phi_l+\cc
\eeq
But if $k\ne l$ this necessarily means that the fields $\phi_k$ and
$\phi_l$ possess the same R-charge. Working out the eigenvalues
of the thermal mass matrix for $\phi_k$ and $\phi_l$ we conclude
that one of the modes has positive thermal mass while the second is
exactly massless. If this happens the fate of the massless mode depends
on its zero-temperature mass.
\bibliography{thmlit}
\bibliographystyle{apsper}
\end{document}